\documentclass[sigplan,9pt,screen,natbib,nonacm]{acmart}
\pdfoutput=1
\setcopyright{none}

\usepackage{hyphenat}
\usepackage{mdwtab}
\usepackage{mdwlist}
\usepackage{mathenv}
\usepackage{amsmath}
\usepackage{comment}
\usepackage{url}
\usepackage{prooftree}

\usepackage{listings}

\lstdefinestyle{sOcaml}{language=[Objective]Caml,
  morekeywords={effect,perform,locus},
  literate={+}{{$+$}}1 {/}{{$/$}}1 
           {=}{{$=$}}1
           {>}{{$>$}}1 {<}{{$<$}}1
           {<>}{$\not=$}1
           {->}{{$\rightarrow$}}2 {>=}{{$\geq$}}2 {<-}{{$\leftarrow$}}2
           {<=}{{$\leq$}}2
           {<<}{{$\ll$}}2
           {==>}{{$\mapsto$}}2
           {|}{{$\mid$}}1
           {'a}{$\alpha$}1
           {'b}{$\beta$}1
           {'c}{$\gamma$}1
           {'d}{$\delta$}1
           {'w}{$\omega$}1
           {'r}{$\rho$}1
           {'state}{$\sigma$}1
           {fn}{$\lambda$}1
           {:=}{\ensuremath{\mathrel{{:}{=}}}}2
           {...}{\ldots}2
           {empty}{\ensuremath\varnothing}1
           {\#\#\#}{{$\leadsto$}}3
}
\newcommand\coleqeq{\mathrel{\mbox{::=}}}
\newcommand\I{\ensuremath{\mathord{\mathsf{int}}}}
\newcommand\ST{\ensuremath{\mathord{\mathsf{string}}}}

\newcommand{\Tr}[1]{\ensuremath{\lceil #1 \rceil}}
\newcommand{\TC}[1]{\ensuremath{\lfloor #1 \rfloor}}

\newcommand\To{\Rightarrow}
\newcommand{\quant}[2]{\mathopen{\text{\divide\thinmuskip2 \divide\medmuskip2 \divide\thickmuskip2 $#1#2.$}\,}}
\newcommand{\fun}{\quant\lambda}
\newcommand{\Bra}[1]{\texttt{<}#1\texttt{>}}
\newcommand{\Esc}[1]{\texttt{\textasciitilde}#1}

\lstnewenvironment{code}{\lstset{basicstyle={\linespread{1.01}\sffamily\normalsize}}}{}

\lstMakeShortInline[columns=fullflexible]|

\newcommand{\aside}[1]{\ignorespaces}

\begin{document}
\title{MetaOCaml\; Theory and Implementation}

\author{Oleg Kiselyov}
\orcid{0000-0002-2570-2186}
\affiliation{%
  \institution{Tohoku University}
  \country{Japan}}
\email{oleg@okmij.org}

\begin{abstract}
Quasi-quotation (or, code templates) has long been used as a
convenient tool for code generation, commonly implemented as a
pre\hyp processing/translation into code-generation combinators. The original
MetaOCaml was also based on such translation, done post type
checking. BER MetaOCaml employs a significantly different, efficient
(especially in version N114) translation integrated with
type-checking, in the least intrusive way.  This paper presents the
integrated efficient translation for the first time.
\end{abstract}
\maketitle

\section{Introduction}
\label{s:intro}

(BER) MetaOCaml \cite{ber-design,ber-metaocaml} is a superset of OCaml to
generate assuredly well-formed, well-scoped and well-typed code using
code templates, also known as brackets and escapes. For example:
\begin{code}
let eta = fun f -> .<fun x -> .~(f .<x>.)>.
 (* val eta : ('a code -> 'b code) -> ('a -> 'b) code = <fun> *)
\end{code}
Although the function looks banal, it has a long history and special
significance in partial evaluation, where it is called `the trick'
\cite{danvy-eta-expansion}.
Brackets |.<...>.| enclose code to generate: in our case, the code of
a function. An escape |.~| marks a hole in the template; the escaped
expression is to generate the code to plug into the hole. Brackets are
akin to string quotation marks `|"|': indeed, a template without holes
can be converted to a string and written into a file. Unlike strings,
however, code templates have structure: the code within a template has to be
well-formed~-- moreover, well-typed OCaml code. If that code has type
|'a|, the whole template has the type |'a code|. Code templates
without holes are values and can be passed as arguments 
(as seen in |f .<x>.|) and returned
as function results. Templates may contain open code, such as
|.<x>.|, which is literally the code of a free variable. Here is an
example of using |eta|, with the detailed reduction sequence:
\begin{code}
eta (fun z -> .<4 * 5 * .~z>.)
$\leadsto_{\beta v}$  .<fun x -> .~((fun z -> .<4 * 5 * .~z>.) .<x>.)>.
$\leadsto_{\beta v}$ .<fun x -> .~(.<4 * 5 * .~(.<x>.)>.)>.
$\leadsto_{splice}$ .<fun x -> .~(.<4 * 5 * x>.)>.
$\leadsto_{splice}$ .<fun x -> 4 * 5 * x>.
\end{code}
The `quoted' (i.e., templated) code remains as is: for example, 
|4 * 5| is not reduced. 
The evaluation~-- substitution of values for bound variables
$\leadsto_{\beta v}$ and filling-in a hole in the template with the 
bracketed value $\leadsto_{splice}$~--
occurs either outside of brackets or within escapes. 
If we enter |eta (fun z -> .<4 * 5 * .~z>.)| at the MetaOCaml top-level,
we indeed see
\begin{code}
- : (int -> int) code = .<fun x_1 -> 4 * 5 * x_1>.
\end{code}
The bound variables get automatically renamed: looking a bit ahead,
choosing fresh variable names is the responsibility of the |mkl| code
combinator, Fig. \ref{f:code-comb}. 

A template without holes and
free variables such as above (so-called close code value) is the
generated code: it can be
written into a file and compiled, and even linked back into the
generated program and invoked there. The function |Runcode.run|
provided by MetaOCaml does the compilation-linking steps:
\begin{code}
let g = Runcode.run .<fun x_1 -> 4 * 5 * x_1>.;;
(* val g : int -> int = <fun> *)
g 3;;
(* - : int = 60 *)
\end{code}
Thus the product |4 * 5| is computed only when the generated code is
compiled and then executed~-- at a later, future stage, so to speak.

(One may get an inkling why |eta| is called `the trick'.)
MetaOCaml is hence a multi-staged language.

This paper presents the theory of BER MetaOCaml implementation.  It
uses the standard in theoretical CS mathematical notation and looks
theoretical. The notation, however, is the \emph{pseudo-code of the
  actual implementation}. The paper is written to prototype in the
mathematical notation the new efficient translation,
\S\ref{s:optimized}, and clarify its subtle points. It is incorporated
into the recently released (May 2023) version N114 of BER MetaOCaml.
The characteristic and surprising feature of the translation is
using what feels like only two stages to
support multiple.

The author could not believe that this is correct, and hence this
paper was written to convince him. The implementation in BER MetaOCaml N114 
was then done by literally transcribing the pseudo-code of
Fig.~\ref{f:optimized} into OCaml. It worked the first time, passing
all tests in the extensive MetaOCaml testing suite.

\section{Type-checking staged programs}

\begin{figure}[htbp]
\begin{tabular}[Ct]{lMl}
Variables & f,x,y,z \\
Types & t \coleqeq \I \mid t\to t
\\
Integer constants & i \coleqeq 0, 1, \ldots
\\
Expressions &
e \coleqeq i \mid x \mid e\; e \mid \fun{x}e
\\
Environment &
\Gamma \coleqeq \cdot \mid \Gamma, x{:}t
\end{tabular}
\caption{Base calculus: simply-typed lambda calculus with integers}
\label{f:Base}
\end{figure}

We start with the base calculus: it is the utterly standard simply typed
lambda calculus with integers, shown merely for the
sake of notation, particularly the notation of the typing judgment:
$\Gamma \vdash e\To e:t$. The notation makes it explicit that type checking
is type reconstruction: converting an `untyped' expression
$e$ to the type-annotated form $e:t$~-- or, in terms of the OCaml type
checker, converting from |Parsetree| to |Typedtree|. \S\ref{s:optimized} shows a
non-trivial use of this notation.

\medskip
\noindent
\begin{prooftree}
\justifies
\Gamma \vdash i\To i:\I
\end{prooftree}
\quad
\begin{prooftree}
x:t \in \Gamma
\justifies
\Gamma \vdash x \To x:t
\end{prooftree}
\quad
\begin{prooftree}
\Gamma \vdash e\To e:t'\to t \quad 
\Gamma \vdash e'\To e':t'
\justifies
\Gamma \vdash e\: e' \To (e:(t'\to t)\; e':t'):t
\end{prooftree}
\\[2\jot]
\begin{prooftree}
\Gamma, x:t' \vdash e\To e:t 
\justifies
\Gamma \vdash \fun{x}e \To (\fun{x:t'} e:t):(t'\to t)
\end{prooftree}
\medskip
\noindent

We assume that the initial environment $\Gamma_i$ to type check the
whole program contains the bindings
of standard library functions such as |succ|, addition, etc.
In the rule for abstraction, one may wonder where does the type $t'$
come from. For the purpose of the present paper, one may consider it
a `guess'. After all, our subject is not type inference, but
staging~-- to which we now turn.

Figure~\ref{f:Staged} presents the staged calculus: the Base
calculus extended with bracket $\Bra e$ and escape $\Esc e$ expression forms
and code types $\Bra t$.

\begin{figure}[htbp]
\begin{tabular}[Ct]{lMl}
Variables & f,x,y,z \\
Types & t \coleqeq \I \mid t\to t \mid \Bra t
\\
Integer constants & i \coleqeq 0, 1, \ldots
\\
Expressions &
e \coleqeq i \mid x \mid e\; e \mid \fun{x}e 
  \mid \Bra e \mid \Esc e
\\
Stage &
n,m \ge 0
\\
Environment &
\Gamma \coleqeq \cdot \mid \Gamma, x^n:t
\end{tabular}
\caption{Staged calculus}
\label{f:Staged}
\end{figure}

The calculus is actually \emph{multi-staged}: brackets may nest
arbitrarily, e.g., $\Bra{\Bra 1}$. The level of nesting
is called \emph{stage}. The present stage, stage 0, is outside of any
brackets. An expression at stage 1 or higher is called future-stage.
As should be clear from the |eta| example in \S\ref{s:intro}, the
evaluation only happens at the present stage.
The typing judgment $\Gamma \vdash_n e\To e:t$ is
now annotated with stage $n\ge 0$. All variable bindings in
$\Gamma$ are also annotated with their stage: $x^n:t$. 

The rules for integer constants and application remain the same,
modulo replacing $\vdash$ with $\vdash_n$: in general, most typing
rules are unaffected by (or, are invariant of) staging. This is a good
news for implementation: adding staging to an extant language does not
affect the type checker to large extent. Here are the changed and new
rules:

\medskip
\noindent
\begin{prooftree}
x^m:t \in \Gamma
\justifies
\Gamma \vdash_n x \To x^m:t
\using{m\le n}
\end{prooftree}
\quad
\begin{prooftree}
\Gamma, x^n:t' \vdash_n e\To e:t 
\justifies
\Gamma \vdash_n \fun{x}e \To (\fun{x^n:t'} e:t):(t'\to t)
\end{prooftree}
\\[2\jot]
\begin{prooftree}
\Gamma \vdash_{n+1} e\To e:t
\justifies
\Gamma \vdash_n \Bra e\To \Bra{e:t}:\Bra t
\end{prooftree}
\qquad
\begin{prooftree}
\Gamma \vdash_n e\To e:\Bra t
\justifies
\Gamma \vdash_{n+1} \Esc e\To \Esc (e:\Bra t):t
\end{prooftree}
\medskip
\noindent

The type-checker also annotates variable references with the stage, in
addition to the type. A variable bound at stage $n$ may be used at the
same stage~-- or higher (but not lower!).  A
present-stage variable may appear within brackets: so-called
\emph{cross-stage persistence} (or, CSP).
As one may expect, bracket increments the stage for its containing
expression and escape decrements. Furthermore, escapes must appear
within a bracket.

For example, 
$\Bra{\Bra{\Esc(\Bra 1)}}$ has the type $\Bra{\Bra\I}$,
the expression $\Bra{\Bra{\fun{x} \Esc(f\: x)}}$ is ill-typed
but
$\Bra{\Bra{\fun{x} \Esc(f\: \Bra x)}}$ is well-typed in an
environment where $f$ is bound to a function $\Bra\I \to \Bra\I$
at stage 0. \S\ref{s:intro} has more examples.

\section{Translating brackets and escapes away}
\label{s:translation}

After a program is type-checked and converted to the type-annotated
form (a.k.a., |Typedtree|), we have to compile it. The type-annotated
form contains brackets and escapes, so our compilation has to account
for them. One popular approach
\cite{calcagno-implementing,chen-meta-programming} is to post-process
the type-annotated expression to eliminate all brackets and
escapes. The post-processed |Typedtree| then has the same form as in the
ordinary OCaml; therefore, we can use the OCaml back-end (optimizer
and code generator) as it is~-- which is what MetaOCaml does.

Formally, the result of post-processing is the Base calculus enriched
with code types (as well as string types and literals) and whose
initial environment contains the functions in
Fig.~\ref{f:code-comb}. We call this calculus Base$_1$.

The post-processing is actually a family of translations: $\Tr{e}$ and
$\Tr{e}^1_n$, which take a type-annotated expression $e:t$ of
Staged calculus and produce the type-annotated Base$_1$ calculus
expression $e'$:
\begin{equation}
\label{e:tran-type}
\Tr{e:t} = e':t \qquad \Tr{e:t}^1_n = e':\Bra t
\end{equation}
The expression $e:t$ in
$\Tr{e:t}$ is a present-stage expression, whereas in
$\Tr{e}^1_n$, it is a $n+1$-stage expression.
This post-processing (optimized in version N102) 
was employed in BER MetaOCaml until the present version N114.

The translation $\Tr{e:t}$ is the identity, until it comes to
bracket:
\begin{eqnarray*}[x]
\Tr{i:\I} = i:\I \quad
\Tr{x^0:t} = x:t  \quad
\Tr{(e\: e'):t} = (\Tr{e}\: \Tr{e'}):t 
\\
\Tr{(\fun{x^0:t'} e:t) : t'\to t} =
(\fun{x:t'} \Tr{e:t}) : t'\to t
\end{eqnarray*}
Switch-over:
\begin{eqnarray*}[x]
\Tr{\Bra{e: t}} = \Tr{e: t}^1_0 : \Bra t \qquad
\Tr{\Esc (e:\Bra t)}^1_0 = \Tr{e : \Bra t}
\end{eqnarray*}
A future-stage translation:
\begin{eqnarray*}[x]
\Tr{i:\I}^1_n = \mathsf{lift_{\I}}\: i : \Bra\I\\
\Tr{x^{m+1}:t}^1_n = x:\Bra t\; (m\le n)\\
\Tr{x^0:t}^1_n = \begin{smcases}
\mathsf{mkid}_t\: \mathsf{"x"}\: : \Bra t & if $x\in \Gamma_i$ \\
\mathsf{lift}_t\: x\: : \Bra t  & otherwise
\end{smcases}\\
\Tr{(e\: e'):t}^1_n = \mathsf{mka}\: \Tr{e}^1_n\: \Tr{e'}^1_n :\Bra t \\
\Tr{\fun{x^{n+1}:t'} e:t}^1_n = 
\mathsf{mkl}\: (\fun{x:\Bra {t'}} \Tr{e:t}^1_n): \Bra{t'\to  t}
\\
\Tr{\Bra{e:t}}^1_n = \mathsf{mkbr}\: \Tr{e:t}^1_{n+1} : \Bra{\Bra t}\\
\Tr{\Esc(e:\Bra t)}^1_{n+1} = \mathsf{mkes}\: \Tr{e:\Bra t}^1_{n} : \Bra t\\
\end{eqnarray*}

\begin{figure}[htbp]
\begin{tabular}[Ct]{Ml@{\;:\;}Ml}
\mathsf{lift}_t & t\to\Bra t \\
\mathsf{mkid}_t & \ST\to\Bra t \\
\mathsf{mka} & \Bra{t_2\to t_1} \to \Bra{t_2} \to \Bra{t_1} \\
\mathsf{mkl} & (\Bra{t_2} \to \Bra{t_1}) \to \Bra{t_2\to t_1} \\
\mathsf{mkbr} & \Bra{t} \to \Bra{\Bra{t}} \\
\mathsf{mkes} & \Bra{\Bra{t}} \to \Bra{t} 
\end{tabular}
\caption{Code-generating combinators}
\label{f:code-comb}
\end{figure}

Figure \ref{f:code-comb} lists the code-generating functions: the producers of
values of the code type.\footnote{In MetaOCaml, they are called
\textsf{lift\_constant\_int}, \ldots,
\textsf{build\_fun}, \textsf{build\_apply}, etc.}
Here $\mathsf{lift}_t$ is the family indexed by type $t$.\footnote{
If such lifting functions exist for all types and how they can be
implemented is a fascinating question that we do not
have space to answer.}

Type preservation of the translation does not seem obvious: after all,
identifiers at any future stage are translated as present-stage
identifiers, of the same name but at the \emph{changed type}: $\Bra
t$, which is furthermore independent of stage. Likewise, functions at a
future stage are translated into present-stage functions, but at a
different type. We discuss the formal properties in in the next
section.  At present we note that all translation equations satisfy
\eqref{e:tran-type}.

For example, the (specialized) |eta| from \S\ref{s:intro}:
\[
\fun{f^0:\Bra\I\to\Bra\I} \Bra{\fun{x^1:\I} \Esc{(f\ \Bra{x})}}
\]
of the type $(\Bra\I\to\Bra\I)\to\Bra{\I\to\I}$
is translated to the Base$_1$ expression
\[
\fun{f:\Bra\I\to\Bra\I} 
\mathsf{mkl}\, \fun{x:\Bra\I} f\, x
\]
clearly of the same type. The origin of the name |eta| should be also clear.

As a more interesting example, consider
\[
\Bra{\fun{x^1:\I} \Bra{\fun{y^2:\I} x + y}}:\Bra{\I\to \Bra{\I\to\I}}
\]
which has two CSPs, both appearing at stage 2: one is $x$, defined at stage 1,
and the other is addition, defined in the initial environment, at stage 0.
The translated Base$_1$ expression is:
\[
\mathsf{mkl}\, \fun{x:\Bra\I} 
\mathsf{mkbr}\,(\mathsf{mkl}\,\fun{y:\Bra\I}
 \mathsf{mka}\, (\mathsf{mka}\, (\mathsf{mkid}\: \textsf{"{+}"})\, x)\, y)
\]
It has the same $\Bra{\I\to \Bra{\I\to\I}}$ type, as one can easily verify.

\subsection{Optimized translation}
\label{s:optimized}

A careful look at the translation rules just presented shows many
opportunities for optimization. First of all, since $\Tr{-}$ is mostly
the identity, it is tempting to cut it out and hence eliminate the
useless traversing and rebuilding of the |Typedtree|. Furthermore,
$\Tr{-}^1_n$ does not essentially use $n$ and can be simplified.

We now present the optimized translation. To avoid $\Tr{-}$ it
requires the integration with the type checker. In principle, we can
combine the translation and the type reconstruction completely.  For
example, the type reconstruction judgments for integer literals would
then become:

\begin{prooftree}
\justifies
\Gamma \vdash_0 i\To i:\I
\end{prooftree}
\qquad
\begin{prooftree}
\justifies
\Gamma \vdash_{n+1} i\To \mathsf{lift_{\I}}\: i : \Bra\I
\end{prooftree}

\medskip
\noindent
That would unwise, however: we have to effectively duplicate the type
checking rules, for stage $0$ and stage $>0$. 
A better idea is to leave the stage-invariant rules (which is most of
them) as
they are and introduce a selective translation
$\TC{e:t}$, defined as the simplified $\Tr{e:t}^1_0$, to
wit:\footnote{performed by \textsf{trx\_translate} of
\url{typing/trx.ml}}
\begin{eqnarray*}[rl]
\TC{i:\I} &= \mathsf{lift_{\I}}\: i : \Bra\I\\
\TC{x^{m+1}:t} &= x:\Bra t\\
\TC{x^0:t} &= \begin{smcases}
\mathsf{mkid}_t\: \mathsf{"x"}\: : \Bra t & if $x\in \Gamma_i$ \\
\mathsf{lift}_t\: x\: : \Bra t  & otherwise
\end{smcases}\\
\TC{(e\: e'):t} &= \mathsf{mka}\: \TC{e}\: \TC{e'} :\Bra t \\
\TC{\fun{x^{n+1}:t'} e:t} &= 
\mathsf{mkl}\: (\fun{x:\Bra {t'}} \TC{e:t}): \Bra{t'\to  t}
\\
\TC{\Esc(e:\Bra t)} &= e:\Bra t\\
\end{eqnarray*}

The typing judgment is now $\Gamma \vdash_n e\To e':t$ where $e$ is
an (un-annotated) expression of the Staged calculus and $e'$ is the
type-annotated expression of Base$_1$ extended with $\Esc{e}$ and
stage-annotated variables. (Bindings in $\Gamma$ are also
stage-annotated. For present stage, the annotation may be dropped.)
Such extended calculus is called Base$_2$. Quite unexpectedly,
Base$_2$ has no need for brackets; it only needs escapes, 
hence the changes to the OCaml |Typedtree| are
minimal. In fact, there are no changes at all, thanks to |Typedtree|
attributes: an escape is indicated by a dedicated attribute attached
to a |Typedtree| node.

\begin{figure}[htbp]
\begin{prooftree}
\justifies
\Gamma \vdash_n i\To i:\I
\end{prooftree}
\qquad
\begin{prooftree}
x^m:t \in \Gamma
\justifies
\Gamma \vdash_n x \To x^m:t
\using{m\le n}
\end{prooftree}
\\[2\jot]
\begin{prooftree}
\Gamma \vdash_n e\To e:t'\to t \quad 
\Gamma \vdash_n e'\To e':t'
\justifies
\Gamma \vdash_n e\: e' \To (e:(t'\to t)\; e':t'):t
\end{prooftree}
\\[2\jot]
\begin{prooftree}
\Gamma, x^n:t' \vdash_n e\To e:t 
\justifies
\Gamma \vdash_n \fun{x}e \To (\fun{x^n:t'} e:t):(t'\to t)
\end{prooftree}
\\[2\jot]
\begin{prooftree}
\Gamma \vdash_1 e\To e:t
\justifies
\Gamma \vdash_0 \Bra e\To \TC{e:t}:\Bra t
\end{prooftree}
\quad
\begin{prooftree}
\Gamma \vdash_{n+2} e\To e:t
\justifies
\Gamma \vdash_{n+1} \Bra e\To 
\Esc (\mathsf{mkbr}\: \TC{e:t}):\Bra t
\end{prooftree}
\\[2\jot]
\begin{prooftree}
\Gamma \vdash_0 e\To e:\Bra t
\justifies
\Gamma \vdash_1 \Esc e\To \Esc (e:\Bra t):t
\end{prooftree}
\qquad
\begin{prooftree}
\Gamma \vdash_{n+1} e\To e:\Bra t
\justifies
\Gamma \vdash_{n+2} \Esc e\To 
\Esc (\mathsf{mkes}\: \TC{e:\Bra t}):t
\end{prooftree}
\caption{Type-checking and translation of Staged into Base$_2$.
}
\label{f:optimized}
\end{figure}

Figure~\ref{f:optimized} presents the pseudo-code of the optimized
translation integrated with type reconstruction. The figure makes it
clear how the Base type reconstruction~-- that is, the |Typedtree|
construction in the ordinary OCaml~-- has to be modified for staging.
Most of the rules (see constant and application rules) are unmodified.
We still need to maintain the stage (as a global mutable variable in
the current implementation). The rule for lambda (and other binding
forms) has to annotate the bound variable with its stage as it is put
into the environment. We do it by adding an attribute bearing the
stage to the |value_description| of the variable. 
The variable rule has to check that
the stage of the variable is less than or equal the current stage, and
to put the stage-annotated variable into |Typedtree|. In the
implementation, nothing needs to be done for the latter: The
|Texp_ident| node of the |Typedtree| carries the |value_description|
taken from the environment, which already has the stage attribute.
The only significant changes are the rules for brackets and escapes
(represented in |Parsetree| as extension nodes).

The selective translation $\TC{-}$ is indeed done only on the parts of
the overall |Typedtree| that represent future-stage sub-expressions.
Therefore, when compiling plain OCaml programs, MetaOCaml imposes \emph{no}
overhead. 

\paragraph{Proposition} 
If $\Gamma \vdash_n e\To e:t$ in the Staged calculus then
$\Gamma \vdash_n e\To e':t$ in the optimized translation.

\paragraph{Proposition} 
If $\Gamma \vdash_n e \To e':t$, then $e'$ has no
  nested escapes.

\paragraph{Corollary}
If $\Gamma_i \vdash_0 e\To e':t$ than $e'$ is strictly a Base$_1$
expression: it contains no escape nodes or stage-annotated
bindings. The type reconstruction hence gives the ordinary OCaml
|Typedtree|, which can then be processed by the OCaml back end as is.

\paragraph{Theorem}
If $\Gamma \vdash_0 e\To e':t$ then 
$\Gamma \vdash \bar{e'}\To e':t$ in Base$_1$
where $\bar{e'}$ is $e'$ with all type annotations removed.

\section{Related work}
\label{s:related}

The idea of implementing code templates by a translation into code
combinators can be traced back to Lisp:
quasi-quotes are commonly implemented as macros, expanding into
S-expression combinators (|cons| and |list|).

The translation  $\Tr{-}$ and $\Tr{-}^1_n$ was implemented in BER
MetaOCaml N101 described in \cite{ber-design}. The translation however
was not presented formally. It is similar to 
\cite[Figure 3]{calcagno-implementing}. However, our
translation is typed. Mainly, we use code combinators instead of data
types and may hence keep the |code| representation abstract.  The
biggest difference is the translation of functions (and other binding
forms such as let expressions, pattern matching and for-loops). We do
not use |gensym|, employing higher-order abstract syntax instead: we
translate a future-stage function also into a present-stage function,
but of a different type, which is then passed to the combinator
\textsf{mkl}. The combinator, among other things, enforces the region
discipline for future-stage variables and checks for scope extrusion,
as described in \cite{ber-design}. 

The optimized translation in \S\ref{s:optimized} is novel: the present
paper is the first presentation of it~-- and BER MetaOCaml N114 is the
first implementation.

\section{Need multiple stages?}
\label{s:disc}

One may have noticed that the eta-generator-generator |etah| in
\S\ref{s:intro} was rather contrived. That is no accident: there are
hardly any realistic examples of needing more than one future stage.
This has been noticed before. In his retrospective
\cite{sheard-accomplishments}, Sheard writes: ``There is no limit to
the number of stages in a MetaML program.  This has been useful
theoretically, but has found very little practical use.  Programmers
find it hard to write programs with more than a few stages.''
\cite[\S21]{sheard-accomplishments}

The only somewhat realistic case I am aware of is generating code that
includes a run-time specializer (evoking just-in-time compilation):
for example, generating code for |power n x| simultaneously with the
code to specialize the power function to a specific value of |n|, and
the overall driver that switches to the specialized version if 
|power n x| was invoked for a specific |n| often enough. (This example was
suggested by Sven Bodo Sholz.)  Even then, such an example seems
better implemented using the tagless-final approach coupled with 
one-future-stage staging.

I would like to ask the readers if there is a value in continuing to
maintain the ability to nest brackets arbitrarily. If not, it would 
make sense to limit the bracket nesting to the single level, which notably
simplifies the implementation.\footnote{One may still build 
generators of generators, using CSP.}

\bibliographystyle{plainnat}
\bibliography{../metafx.bib}

\providecommand{\leanTAP}{\mbox{{\sf lean}\smash{$T^{\!\!\textstyle A}\!\!P$}}}
\begin{thebibliography}{6}
\providecommand{\natexlab}[1]{#1}
\providecommand{\url}[1]{\texttt{#1}}
\expandafter\ifx\csname urlstyle\endcsname\relax
  \providecommand{\doi}[1]{doi: #1}\else
  \providecommand{\doi}{doi: \begingroup \urlstyle{rm}\Url}\fi

\bibitem[Calcagno et~al.(2003)Calcagno, Taha, Huang, and
  Leroy]{calcagno-implementing}
Cristiano Calcagno, Walid Taha, Liwen Huang, and Xavier Leroy.
\newblock Implementing multi-stage languages using {AST}s, gensym, and
  reflection.
\newblock In \emph{{GPCE}}, number 2830 in {L}ecture {N}otes in {C}omputer
  {S}cience, pages 57--76, 22--25 September 2003.
\newblock \doi{10.1007/978-3-540-39815-8_4}.

\bibitem[Chen and Xi(2005)]{chen-meta-programming}
Chiyan Chen and Hongwei Xi.
\newblock Meta-programming through typeful code representation.
\newblock \emph{Journal of Functional Programming}, 15\penalty0 (6):\penalty0
  797--835, 2005.
\newblock \doi{10.1017/S0956796805005617}.

\bibitem[Danvy et~al.(1996)Danvy, Malmkj{\ae}r, and
  Palsberg]{danvy-eta-expansion}
Olivier Danvy, Karoline Malmkj{\ae}r, and Jens Palsberg.
\newblock Eta-expansion does {T}he {T}rick.
\newblock \emph{{ACM} Transactions on Programming Languages and Systems},
  18\penalty0 (6):\penalty0 730--751, 1996.

\bibitem[Kiselyov(2014)]{ber-design}
Oleg Kiselyov.
\newblock The design and implementation of {BER} {MetaOCaml} - system
  description.
\newblock In \emph{{FLOPS}}, number 8475 in {L}ecture {N}otes in {C}omputer
  {S}cience, pages 86--102. Springer, 2014.
\newblock \doi{10.1007/978-3-319-07151-0\_6}.

\bibitem[Kiselyov(2023)]{ber-metaocaml}
Oleg Kiselyov.
\newblock {BER MetaOCaml N114}.
\newblock \url{https://okmij.org/ftp/ML/MetaOCaml.html}, May 2023.

\bibitem[Sheard(2001)]{sheard-accomplishments}
Tim Sheard.
\newblock Accomplishments and research challenges in meta-programming.
\newblock In Walid Taha, editor, \emph{Proceedings of {SAIG} 2001: 2nd
  International Workshop on Semantics, Applications, and Implementation of
  Program Generation}, number 2196 in {L}ecture {N}otes in {C}omputer
  {S}cience, pages 2--44, Berlin, 6~September 2001. Springer-Verlag.
\newblock ISBN 3-540-42558-6.

\end{thebibliography}
\end{document}